\documentclass[manuscript, 11pt]{acmart}
\usepackage{subcaption}
\usepackage{placeins}  
\usepackage{float}

\author{Mingyue Zha}
\affiliation{%
  \institution{Dartmouth College}
  \city{Hanover}
  \country{US}}
\email{Mingyue.Zha.27@dartmouth.edu}

\author{Ho-Chun Herbert Chang}
\affiliation{%
  \institution{Dartmouth College}
  \city{Hanover}
  \country{US}}
\email{Herbert.Chang@dartmouth.edu}

\AtBeginDocument{%
  }

\begin{document}

\title{Gender Inequalities in Content Collaborations:
Asymmetric Creator Synergy and Symmetric Audience Biases}

\begin{abstract}
Content-creator collaborations are a widespread strategy for enhancing digital viewership and revenue. While existing research has explored the efficacy of collaborations, few have looked at inequities in collaborations, particularly from the perspective of the supply and demand of attention. Leveraging 42,376 videos and 6,117,441 comments from YouTube (across 150 channels and 3 games), this study examines gender inequality in collaborative environments. Utilizing Shapley value, a tool from cooperative game theory, results reveal dominant in-group collaborations based on in-game affordances. However, audience responses are aligned across games, reflecting symmetric biases across the gaming communities, with comments focusing more on peripherals than actual gameplay for women. We find supply-side asymmetries exist along with demand-side symmetries. Our results engage with the larger literature on digital and online biases, highlighting how genre and affordances moderate gendered collaboration, the direction of inequality, and contributing a general framework to quantify synergy across collaborations. 
\end{abstract}



\maketitle

\section{Introduction}
Gaming communities are steadily gaining mainstream popularity and have become a focal point of interest for cooperative work and social computing scholarship~\cite{barr2007video, zakaria2022online}. Video platforms like YouTube and streaming platforms like Twitch have become the loci of vibrant gaming communities, many of which were cultivated during the pandemic. However, while a growing body of literature has addressed streaming~\cite{wu2023interactions, wu2023streamers,chen2024felt}, less have studied the collaboration of  individuals within these communities. 

Broadly, content collaboration, in which two or more content creators come together to co-produce media, is a key strategy on social media platforms to increase audience interaction and engagement~\cite{googlesupport, mediumcollab}. Through these means, content collaborations serve as strategic mechanisms to diversify and cross-promote content across communication networks. Recent empirical study shows that on video sharing sites such as YouTube, approximately 17\% of the video content observed is collaborative content. The growth and efficacy of collaborations has stimulated research and adjustments in platform affordances conducive to collaboration~\cite{koch2018collaborations}. For example, \textit{Bilibili}, a leading video sharing website in China, introduced a content collaboration feature for the community in 2019~\cite{cmbi2020bilibili}. In 2021, Instagram also introduced a feature that allows users to add co-authors to a post, allowing the same post to show up under multiple profiles~\cite{later}. 

In tandem, video games are immersive environments connecting millions of people in virtual worlds~\cite{ESA2024}, where individuals communicate and socialize, facilitating verbal and non-verbal interactions~\cite{li2019measuring, wang2019effect}. These  games and their associated communities are used to fulfill social needs, allowing players to engage with friends, meet people across the world, and gather for community-based events~\cite{rogers2017motivational, hilvert2018social, barreda2022psychological}. However, because gaming environments are embedded within everyday social experiences, these environments often mirror and exacerbate real-world social inequalities~\cite{apperley2020digital, cote2020gaming}. 

While past academic studies have looked at the overall impact of collaborations on channel growth and popularity, and although many studies investigate gender inequalities in game play or streaming directly, few have investigated inequalities in collaborations, as well as robust methods for social scientific measurement. 
In this study, we examine the inequities across genders and quantify synergy in content-creator collaborations. We examine the gender-based collaborations of the top YouTube content creators across three games: \textit{Valorant}, \textit{Animal Crossing}, and \textit{Dead by Daylight}. These games were chosen due to their streaming popularity, high parity for women gamers, and ability to showcase three disparate game genres: first-person-shooter ("FPS") games, cozy games, and horror games.

Our study provides three contributions. First and substantively, we provide evidence regarding gender asymmetries in content collaborations in pure viewership and network dynamics. Collaborations increase viewership, but the level of synergy may skew toward specific pairs of gendered collaborations. By comparing three different games known for their content-creator communities, we evaluate qualitatively how affordances impact collaborations and address an existing gap in cross-platform or cross-game research. Second, we provide evidence from both the supply and demand side of attention: while collaboration dynamics differ based on game affordances, downstream audience discourse is dictated by the gender of collaborating creators. Third and methodologically, we provide a general framework to assess gender biases and collaborative synergy for other online sub-community-based activities, drawing from game theory.

\section{Background}

\subsection{Content-Creator Collaborations}

Content collaborations involve multiple creators coming together to produce, share, or enhance content in digital environments and communication landscapes. Under a dyadic collaborative strategy (the most common form of content collaboration), Channel A, owned by a content creator, will introduce a collaborating Channel B to their viewers to attract Channel B’s community to Channel A’s content and vice versa. Such collaborations often occur reciprocally on both of the channels to attract each other’s viewers and increase both channels’ visibility and fanbases~\cite{pan2023leveraging}. 

Past studies on the impact of content collaborations have found that median view counts are higher for videos featuring collaborations compared to videos with no collaborations as well as a slight positive effect on channel subscribers. For instance, Koch et al. [2018] directly compares viewership and channel/subscriber growth as a result of collaborating and non-collaborating videos~\cite{koch2018collaborations}. Yang et al. [2023] uses a game theory model with creators competing for consumers on a Hotelling line, and shows that collaboration allows creators to use the jointly-produced content to moderate competition, while using their individual content to expand into new audiences~\cite{yang2023collaboration}. Irawan et al. [2023] measured viewer perception of video collaborations using sentiment analysis, showing that different genres of content in collaborations yield differing changes to sentiment values~\cite{irawan2023viewer}. 

Notably, these studies focused on the difference between collaboration and non-collaboration, rather than across collaboration paradigms--such as those defined by gender. However, in collaborations, success and outcomes are often predicated upon social interactions, communication dynamics, and power relations~\cite{dietrich2010dynamics}. 

To fill this gap, we apply game theory and a comparison of collaborative synergies to identify inequalities. Collaborations come in different efficacies; the joint engagement on a co-produced work may be more or less than the sum of their parts. In the behavioral contagion literature, co-diffusion is often defined as synergistic (greater than the sum or product) and antagonistic (less than the sum or product)~\cite{chang2018co,chang2019co}, typically on social networks and focuses on diffusion depth and speed. A similar measure has yet to be proposed for content collaborations, as existing studies mostly focus on direct comparisons between the outcomes of collaborating versus non-collaborating content. We propose a novel application of the Shapley value to quantify contributions each channel makes towards the outcome of the collaboration. 

We also apply affordance theory to explain collaboration differences between genres of game content. Affordances are the possibilities of action that someone has within their environment, and technological affordances are useful for examining technology and human relationships~\cite{chemero2018outline}. Volkoff and Strong defined affordances as “the potential for behaviors associated with achieving an immediate concrete outcome and arising from the relation between an object and a goal-oriented actor or actors”~\cite{bygstad2016identifying}. Here, affordances in games will vary depending on game mechanics and is simultaneously influenced by player demographics and sociocultural affordances.

\subsection{Gender Inequity in Gaming Communities}

Existing research has provided valuable insights into the causes and manifestation of inequalities and exclusion in gaming. Social scientists have theorized how affordances of different video game environments would change human behavior–specifically, the factor of anonymity within gaming environments~\cite{kordyaka2020towards} and the zero-sum designs inherent to many games~\cite{adinolf2018toxic}. 

First, the Social Identity Model of Deindividuation Effects (SIDE) theory describes how in information-sparse settings (i.e. anonymous environments), individuals do not lose their sense of identity but shift from personal to social identity, aligning behavior with group norms~\cite{lea2001knowing}. However, this also means group-based biases can be exacerbated by viewers. Often when female gamers reveal their gender identities online, gamers tend to cease speaking about game-related topics and instead shift the conversation to the gamer herself~\cite{su2011virtual,nardi2010my,nakandala2017gendered}. This may manifest strongly in commentors (demand-side) where their identities are obscured. 

Second, the zero-sum nature of many games promote exclusion and hostility towards certain players while rewarding and granting privileges to players who perform better~\cite{paul2018toxic}. These dynamics ignore players' access to and level of engagement with games, which can depend on factors outside of their control, such as time invested, support from peers, and resources put towards games. Meritocratic ideals are often weaponized against disadvantaged groups in gaming, reproducing and perpetuating social inequality~\cite{massanari2017gamergate}. One way this manifests is in the presumption that online "hardcore" gaming communities are predominantly male with little discernible feminine influence~\cite{richard2018gendered}. And although women are underrepresented and marginalized across the gaming industry, from gaming-related events, major league gaming, and content creation, these patterns are often written off as an indicator of lack of competence in the related spheres of gaming~\cite{prescott2011segregation, darvin2021breaking}. 

We formalize existing gender biases in gaming communities and apply them to content collaborations from two directions: supply-side inequalities and demand-side biases. This is a common framework in social media studies that conceptualize online ecosystems as a combination of content creators, their audiences, and the algorithms that mediate them~\cite{munger2022right}. In the context of streaming, the supply-side is the content creators and their collaborations; the demand-side is the audience interactions. What this paradigm does not consider is that fact that beyond the streaming platform, the games themselves and their affordances do much to mediate the content creators and audience. However, differences in these affordances are largely qualitative in nature, which motivates a case-study approach to understanding how affordances may dictate biases.

\subsection{Three Case Studies}

Games often have highly segmented demographics such as gender, which was noted qualitatively as early as the 1990's between \textit{Mortal Kombat} versus \textit{Barbie}, or more recently using computational approaches in games such as those in the \textit{otome} genre that appeal to young women~\cite{cassell2000barbie, lei2024game}. For this paper, we chose three games that are well-known for robust streaming communities that span genres and have a known, gendered dimension.

First, \textit{Valorant} is an FPS game that emphasizes teamwork, strategy, and action and has a competitive community with high engagement in esports and streaming. Not only is it the most-watched game on Twitch~\cite{thespike}, but it is also ideal for studying gendered collaboration dynamics. While it is a male-dominate game, it has a significantly higher female ratio, with its 30-40\% female player base, compared to other FPS games at 4-7\%~\cite{mediumvalorant, cnnvalorant}. Due to its relative gender parity, there are more opportunities for gender comparisons in \textit{Valorant} than other FPS games. 

Second, \textit{Animal Crossing} is a simulation game with a focus on creativity, community building, and social interaction. Its relaxed gameplay and the way creators share content by visiting each other's islands provide insights into more casual, creative collaborations~\cite{ignac}. It is one of if not the most well-known life simulation games and has seen enduring popularity since the start of its franchise in 2001~\cite{theliberator}. \textit{Animal Crossing: New Horizons}'s release in 2020, coinciding with the pandemic, led to even greater virtual engagement~\cite{pennstate}. Moreover, \textit{Animal Crossing} is a rare game with an almost even split between genders, slightly favoring women gamers~\cite{yougov, thegamerac}.

Third, \textit{Dead by Daylight} is one of the most popular horror games. In 2022, \textit{Dead by Daylight} surpassed 50 million players globally~\cite{kitguru}. Within the game, there is an asymmetrical multiplayer experience where one player is a killer and others are survivors. This requires cooperation among survivors and strategic thinking from the killer, leading to unique collaboration challenges~\cite{thegamerdbd}. Moreover, it is regarded as having one of the greatest following among LGBTQ+ gamers, which allows us to qualitatively surface distinctions between gender and sexual orientation~\cite{dexerto}.

The chosen games appeal to different demographics and gaming communities, allowing us to analyze collaboration dynamics across various player bases. \textit{Valorant} attracts a competitive demographic; \textit{Animal Crossing} attracts a casual demographic; \textit{Dead by Daylight} falls in between, with a mix of casual and dedicated players~\cite{theportal, theverge, pcgamer}. More crucially, each game presents a relatively dominant gender that may influence the supply and demand dynamics of attention. 

\subsection{Research questions}

Our principal aim is to balance quantitative analyses of gendered collaboration dynamics with qualitative observations of games and their affordances. While a more systematic evaluation across a larger number of games is necessary, our aim is to provide a framework for future investigation more broadly. 

With that in mind, our research questions are as follows:
\begin{itemize}
    \item \textbf{Supply side:} How do collaboration synergies diverge based on gender and gendered dyads a) across games and b) within games? 
    \begin{itemize}
        \item \textbf{H1a}: Gendered in-group ties will generate the greatest synergy.
        \item \textbf{H1b}: Men will be more central in networks than women.
    \end{itemize}
    \item \textbf{Demand side}: How do audience sentiment and topics diverge between content creators when collaborating?
    \begin{itemize}
        \item \textbf{H2a}: Men-led collaborations will elicit more toxic reactions.
        \item \textbf{H2b}: Audience comments towards men will discuss gameplay more than comments towards women.
    \end{itemize}
\end{itemize}

\section{Methods}

\subsection{Data Acquisition and Processing}

Our data consists of YouTube videos from the 50 most popular YouTube content creators for each of the three games. We defined the 50 most popular English-speaking YouTube content creators by their number of subscribers. First, we cross-referenced community rankings from sources like Feedspot and Ranker, which reflect audience perceptions of the most popular channels for each game~\cite{feedspot, ranker}. Thus, these channels are widely recognized for dedicating a majority of their videos to gaming content of \textit{Valorant}, \textit{Animal Crossing}, or \textit{Dead by Daylight}. Then, within those lists of channels, we narrowed down to the top 50 by highest subscriber count. We determined the gender identity of content creators by examining how they self-identified in their social media biographies and the pronouns used by their audience to refer to the creator. All 150 content creators identified within the gender binary.

While there are certainly hundreds of YouTubers for each game, studies have found viewership and compensation for streaming regularly follow a power law--commonly known as the rich get richer phenomenon--where the top YouTubers take in the lion's share of views and revenue~\cite{adamic2000power,yu2015lifecyle, houssard2023monetization}. As we are interested in not just the behavior of streamers, but their interactions with the audience, 50 per game provided a suitable and feasible cut-off for our study. 

We collected metadata for YouTube videos, including titles, descriptions, and engagement metrics, using the YouTube Data API from Google Developers. These metrics are shown in Table~\ref{tab:freq}. Scraping video metadata with the YouTube API, results are constrained to a maximum of 500 videos per channel~\cite{youtubeapi}. In April 2024, we scraped metadata for the most recent 500 videos per channel, covering content uploaded between January 2011 to April 2024, resulting in a total of $n = 42,376$ videos. Additionally, using the video IDs obtained, we scraped all comments below the videos, yielding a total of $n = 6,117,441$ comments. 

\begin{table}[h]
  \caption{Video Distribution and Gender Distribution of Top 50 Gaming YouTubers Across Games}
  \label{tab:freq}
  \begin{tabular}{cccc}
    \toprule
    Game & Videos & \# Men & \# Women \\
    \midrule
    Valorant & 13,471 & 42 & 8\\
    Animal Crossing & 13,367 & 15 & 35\\
    Dead by Daylight & 15,538 & 42 & 8\\
  \bottomrule
  \Description{Table showing the number of videos and gender distribution of the top 50 gaming YouTubers for Valorant, Animal Crossing, and Dead by Daylight. Valorant and Dead by Daylight both have 42 men and 8 women creators, while Animal Crossing has 15 men and 35 women creators. The number of videos ranges from 13,367 for Animal Crossing to 15,538 for Dead by Daylight.}
\end{tabular}
\end{table}

\subsection{Computing Synergy}

To examine collaboration synergy, we use a novel application of the Shapley value analysis from cooperative game theory, which quantifies the contribution of each individual within a coalition~\cite{winter2002shapley}. A collaboration occurs when one YouTuber’s unique user handle appears in another YouTuber’s video description. In each collaboration, two actors are involved: Channel A, the channel that posts and owns the video, and Channel B, the collaborator that is appearing in the video. Using Shapley value analysis, we measure the increase in viewership each player contributes when they collaborate. For instance, the two-way Shapley value (or contribution) of Channel A, $SHAP_2(A)$, is calculated by Eq. ~\ref{eq:shap} where $\tilde{x}(B)$ is the median viewership of Channel B. Shapley values were then aggregated by gender.
\begin{equation} \label{eq:shap}
    SHAP_2(A) = \frac{1}{N} \sum V_i(A+B) - \tilde{x}(B)
\end{equation}
Depending on the genders of the YouTubers of Channel A and Channel B, there are four possible gendered collaborations: Woman-Woman ("W-W"), Woman-Man ("W-M"), Man-Woman ("M-W"), and Man-Man ("M-M"). We indicate the gender of Channel A first. 

We normalize these values since different games have different levels of popularity, and raw Shapley values would be skewed by these differences. For $SHAP_2(A)$, we normalized with Eq.~\ref{eq:shap-norm}:
\begin{equation} \label{eq:shap-norm}
    SHAP_n(A) = \frac{SHAP_2(A)}{\tilde{x}(B)} - 1
\end{equation}
Since $SHAP_2(A)$ determines the benefit Channel A brings to Channel B, we normalize with median views of Channel B. We do the reverse for normalizing $SHAP_2(B)$.

Though video collaborations with more than two creators exist, we only considered two-way collaborations because this allowed the clearest determination of collaboration synergies between the actors. Additionally, within all collaborations, the majority (69.6\%) were two-way collaborations. This is large in context of analyzing collaborations. With 50 streamers, the possible two-way combinations are ${50 \choose 2} = 1,225$; possible three-way combinations are ${50 \choose 3} = 19,600$.

\subsection{Data Augmentation with Deep Learning}

Additionally, we compute the sentiment of audience comments. We labeled sentiment scores for each comment using VADER, a lexicon and rule-based sentiment analysis tool~\cite{elbagir2019twitter}. While sentiment analysis has limitations, comparisons across different collaborative ties using a fixed methodology can still capture meaningful divergences in audience tone. To improve interpretability, we first use BERTopic to identify salient issue topics. BERTopic is a topic modeling technique that leverages BERT embeddings to create clusters of semantically similar documents, providing interpretable topics from large collections of text~\cite{grootendorst2022bertopic}. 

We then applied zero-shot classification to identify key relevant metrics, including gameplay, stream environment, food, person’s appearance, and other. Zero-shot classification is a machine learning approach where a model is trained to classify data into categories that it has never seen before, relying on semantic understanding and contextual clues. This technique is particularly useful in natural language processing, where it leverages pre-trained language models to generalize across different tasks without requiring task-specific training data~\cite{wang2022pre}. In particular, appearance versus gameplay is an important criterion we wish to compare across gendered collaborations.

\subsection{Network Analysis}

Network analysis is a common strategy to understand the structure of interaction within a domain. A network, or graph, is defined by nodes connected by edges. 

We create supply-side and demand-side networks to capture the interactions between content creators and viewers in each gaming community. In the supply-side network, we define content creators as nodes and collaborations as edges. We then compute the closeness centrality for each user and aggregate by gender. Closeness centrality is a measure in network analysis that quantifies how close a node is to all other nodes in the network, based on the average shortest path length from that node to all others~\cite{zhang2017degree}. A node with high closeness centrality in our case indicates they have collaborated closely with other key nodes in the network. Overall, this allows us to understand gender inequality based on their network position. 

In the demand-side network, nodes are content creators and viewers, while edges indicate the frequency of direct interactions, or comments, between individual viewers and creators. In this network, we assess the extent commenters bridge their engagement between different streamers and how diverse viewer engagement in different gaming communities are.

\subsection{Diversity Analysis}

To operationalize diversity, we use the Shannon entropy measure, developed in the biological sciences~\cite{magurran2013ecological,begon2021ecology} but now commonly used in human behavioral science~\cite{gallagher2017disentangling,chetty2022social,chang2023liberals}. In network science, entropy is a key measure of randomness and diversity within a network's structure. Specifically, it quantifies the uncertainty or unpredictability in the distribution of connections between nodes. To assess the diversity of commenters interacting with content creators in each game, we calculated the entropy of the commenters within the network. Shannon entropy $H$ examines the entropy related to the arrangement and connectivity of nodes in the streamer-viewer network. The Shannon entropy of a network is given Eq.~\ref{eq:entropy}:
\begin{equation} \label{eq:entropy}
H = - \sum p(x) log_2 p(x)
\end{equation}
This formula calculates entropy $H$ as the negative sum of the probability $p(x)$ times the logarithm base 2 of that probability. Higher values of $H$ indicate a more diverse distribution of commenter interactions across content creators, implying greater randomness and less concentration in the network.

\section{Results}

\subsection{Supply-Side Analysis: Game Affordances Influence Frequency of Collaborations}

\begin{figure}[ht]
  \centering
  \includegraphics[width=0.6\linewidth]{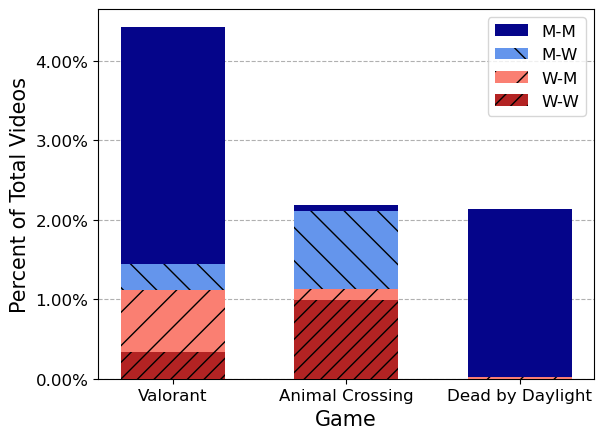}
  \caption{Percentage of Collaborations Over All Videos, by Collaboration Pair with Man-Man (Navy), Man-Woman (Light Blue), Woman-Man (Salmon), and Woman-Woman (Red).}
  \label{fig:dyad-summary}
  \Description{Stacked bar chart showing the gendered distribution of the percentage of collaborating videos over all videos across Valorant, Animal Crossing, and Dead by Daylight. Percentages from 0 to 4.5\% are on the y-axis, the three games are on the x-axis. Each bar is stacked from top to bottom with the percentages of gendered collaboration pairs for Man-Man, Man-Woman, Woman-Man, and Woman-Woman collaborations."}
\end{figure}

First, game affordances dictate the conduciveness of collaborations. Four types of gendered collaborations were noted: Woman-Woman, Woman-Man, Man-Woman, and Man-Man. Figure~\ref{fig:dyad-summary} shows that 4.43\% of \textit{Valorant} videos, 2.18\% of \textit{Animal Crossing videos}, and 2.14\% of \textit{Dead by Daylight} videos are two-set collaborations. 

We posit that the \textit{Valorant} creator community was most conducive to collaborations due to their in-game affordances--collaboration opportunities that are afforded by a game’s core mechanics~\cite{hamalainen2018raise}. For example, \textit{Valorant}’s unique duo-play functionality and particularly competitive nature provide a more conducive environment for content creators to collaborate. This is an extension of previous work which found different YouTube video genres have different affordances for collaborations. Genres such as entertainment, vlogging, and cooking allow greater ease for human interactions and encourage more collaborations between creators~\cite{koch2018collaborations}. 

The gender distribution among collaborations also matters. Looking at gender symmetry, \textit{Valorant} has a similar number of M-W and W-M collaborations, indicating there may be greater reciprocity between collaborators. However, \textit{Animal Crossing} has a much larger share of M-W collaborations compared to W-M collaborations, indicating less reciprocity. In addition, the principal form of collaboration is with women as the secondary streamer. Lastly, Dead-by-Daylight is dominated by M-M pairings, in part due to the high concentration of streamers who are men. 

What these findings show is that a game's genre and audience significantly determine the dynamics of collaboration. \textit{Dead by Daylight}, with its popularity specifically in the LGBTQ+ community, is dominated by M-M collaborations. Even among games with higher woman streamers, FPS games may engender more M-M collaborations, whereas in casual sandbox games like \textit{Animal Crossing}, collaborations ties are usually supported by women.

\subsection{Supply-Side Analysis: Collaboration Synergies Vary Between Games}

\begin{table}[h] 
  \caption{Normalized Shapley Values for Each Channel. a) Shows the Shapley Values, or Contribution, of Channel A. b) Shows the Shapley Values, or Contributions, of Channel B.} \label{tab:shap-values}
  \begin{subtable}{0.45\textwidth}
    \subcaption{Normalized Shapley Values for Channel A}
    \begin{tabular}{lcccc}
    \toprule
    Game & W-W & W-M & M-W & M-M\\
    \midrule
    Valorant & 0.777 & 1.733 & -0.728 & \textbf{4.926}\\
    Animal Crossing & \textbf{8.977} & 2.195 & 0.690 & 0.108\\
    Dead by Daylight & — & -1.727 & \textbf{10.391} & 3.694\\
  \bottomrule
  \end{tabular}
  \end{subtable}

  \vspace{10pt}
  \begin{subtable}{0.45\textwidth}
    \subcaption{Normalized Shapley Values for Channel B}
    \begin{tabular}{lcccc}
    \toprule
    Game & W-W & W-M & M-W & M-M\\
    \midrule
    Valorant & 0.362 & 1.041 & 0.0429 & 0.338\\
    Animal Crossing & 0.0711 & -0.548 & -0.878 & 0.845\\
    Dead by Daylight & — & -1.295 & -1.210 & -0.311\\
  \bottomrule
  \Description{Table showing normalized Shapley values representing the contribution of different games across gender matchups (W-W, W-M, M-W, M-M) across Valorant, Animal Crossing, and Dead by Daylight for Channel A (subtable a) and Channel B (subtable b).}
  \end{tabular}
  \end{subtable}
\end{table}

Our next step is to directly quantify the synergies of these collaborations, which we present in tabular form for within game, across game, and across gender pair comparisons. Table~\ref{tab:shap-values} shows the normalized Shapley values for each of the games. The way to interpret the numbers is as follows: Table~\ref{tab:shap-values}a) values are the amount that Channel A contributes to the relationship, and Table~\ref{tab:shap-values}b) values are the amount that Channel B contributes. The higher the value, the greater the contribution. The highest value of each row indicates the greatest synergy or additive effect across gendered ties for a particular game. The highest value of each column shows the game with the greatest gender synergy relating to that column. 

Looking at Table~\ref{eq:shap-norm}a), within \textit{Valorant}, men streamers tend to benefit more strongly from other men streamers, as demonstrated by the high M-M $SHAP_n(A)$ value. In \textit{Animal Crossing}, women contribute the most value to other women. \textit{Dead by Daylight} has the highest Shapley value for M-W collaborations--which suggests that men YouTubers are most effective in synergizing with women YouTubers in \textit{Dead by Daylight}.

The heterogeneity of these results shows the affordances and dominant player gender of these games may greatly influence the collaboration dynamics. However, each game has a tie with significantly more asymmetry. In more competitive, FPS games like \textit{Valorant}, the Channel A in M-M immensely increases viewership. For games with more traditionally feminine attributes, like \textit{Animal Crossing}, Channel A contributes most in W-W ties. In both cases, the dominant gender of the game generated the highest synergy for the same gender. For games with a large LGBTQ+ community, M-W ties have the highest $SHAP_n(A)$ value. With our limited case study, we find game affordances tend to dominate specific gender-based patterns in collaboration.

\begin{figure}[h] 
  \centering
  \includegraphics[width=.6\linewidth]{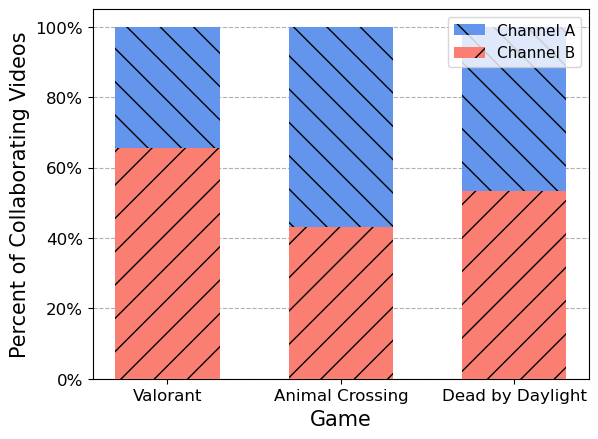}
  \caption{Percentage of Videos Where Channel A (Light Blue) or Channel B (Salmon) Has Greater Median Viewership.}
  \label{fig:reciprocity}
  \Description{Stacked bar chart showing the percentage of instances where Channel A (light blue) or Channel B (salmon) has a greater median viewership in a collaboration set across a) Valorant, b) Animal Crossing, and c) Dead by Daylight. Percentages from 0 to 100\% are on the y-axis, and the three games are on the x-axis. Channel B is the more popular channel for most Valorant and Dead by Daylight collaborations while Channel A is the more popular channel for most collaborations in Animal Crossing.}
\end{figure}

Additionally, when comparing Shapley values for Channel A and Channel B, most of the collaborations have a greater $SHAP_n(A)$ than $SHAP_n(B)$. This indicates that Channel A tends to have an outsized contribution towards the collaboration. Surprisingly, in Figure~\ref{fig:reciprocity}, when looking at whether Channel A or Channel B has more views on average on their non-collaborating videos, the results vary between games. For \textit{Valorant} and \textit{Dead by Daylight}, Channel A tends to be less popular, meaning collaborations flow upstream. For \textit{Animal Crossing}, collaborations tend to flow downstream, with Channel A being more popular. 

We chose to include \textit{Dead by Daylight} as a key focus due to its significant LGBTQ+ player presence, which could provide a unique context to explore the effects of this demographic on gendered collaborations. However, we observed that the top streamers within this community were overwhelmingly men, and there were no W-W collaborations in this dataset. This introduces a limitation to our study, which we will further address in the discussion section.

\subsection{Supply-Side Analysis: Creator Network Centrality}

\begin{figure}[h] 
  \centering
  \includegraphics[width=\linewidth]{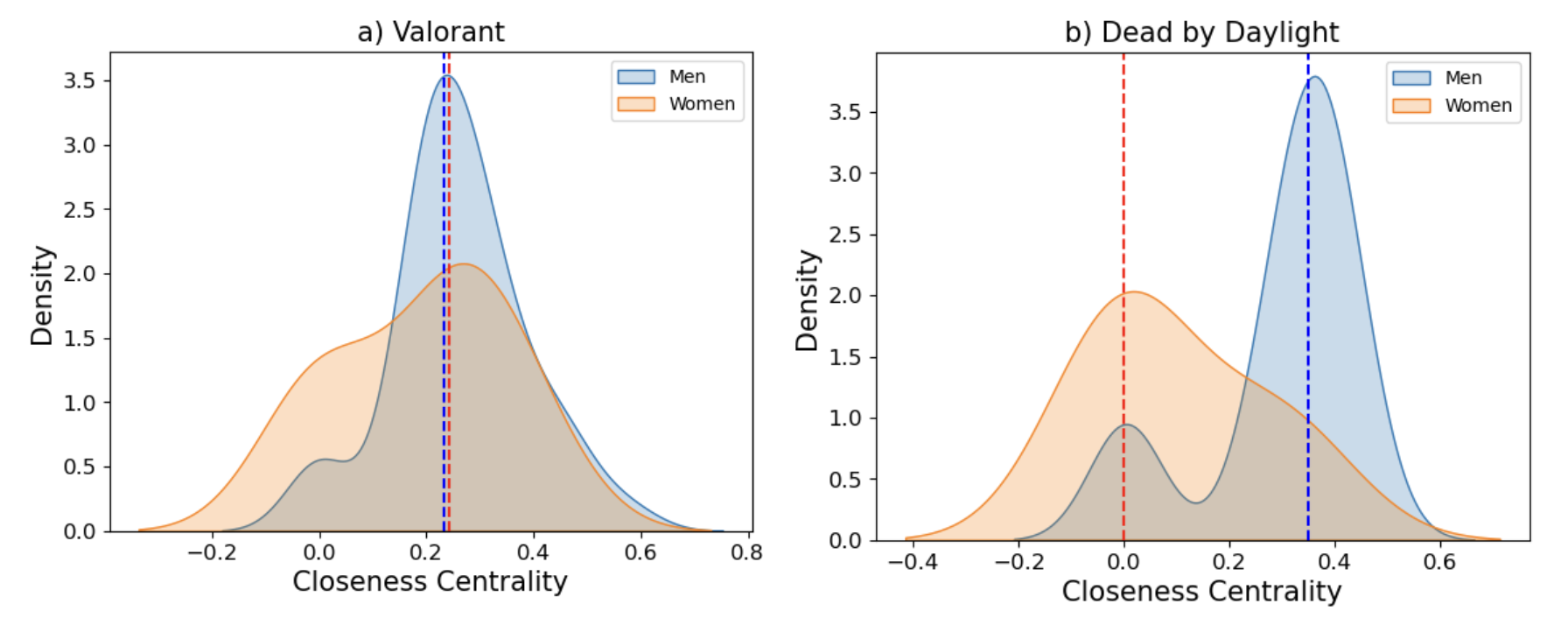}
  \caption{Closeness Centrality of Men (Light Blue) and Women (Orange) YouTubers in a) Valorant and b) Dead by Daylight.} 
  \Description{Density plots comparing the closeness centrality of men (blue) and women (orange) content creators in collaboration networks in a) Valorant and b) Dead by Daylight. Densities ranging from 0 to 3.5 are on the y-axis, and closeness centralities from 0 to 0.8 are on the x-axis. The vertical dashed lines indicate the median closeness centrality for both men (blue dashed line) and women (red dashed line).}
  \label{fig:centrality}
\end{figure}

Another way of exploring collaboration dynamics is by considering network centralities. Figure~\ref{fig:centrality} shows the distribution of closeness centrality for men (blue) and women (orange) in collaboration networks. Closeness centrality is a measure of how close a particular node is to all other nodes (e.g., channels) in the network. 

For \textit{Valorant}, while men hold more positions within the top 50 YouTubers, women have a slightly higher median closeness centrality (0.24 for women versus 0.23 for men). This means in terms of averages, \textit{Valorant} is gender symmetrical in how connected women gamers are in collaboration networks. 

The story is flipped for \textit{Animal Crossing}. Though women hold more positions in the top 50 content creators, men have a median closeness centrality of 0.22 while women have the lower median of 0.19.

Finally, \textit{Dead by Daylight} is most gender asymmetrical in its network centrality. Women have a median closeness centrality of zero while men have a median of 0.36. 

The median centrality for each gender varies across games; however, women creators consistently exhibit flatter density curves, suggesting a higher concentration of women at the peripheries of collaboration networks compared to men. While network centralities provide insight into the collaborative opportunities available to creators, they do not account for the exchange of viewership across network connections. Thus, the application of Shapley value analysis is crucial for examining the synergy between collaborative ties.

\subsection{Demand-Side Analysis: Woman-Led Collaborative Streaming Yields More Positive Tone but Less Discourse on Gameplay}
\begin{figure}[h] 
  \centering
  \includegraphics[width=\linewidth]{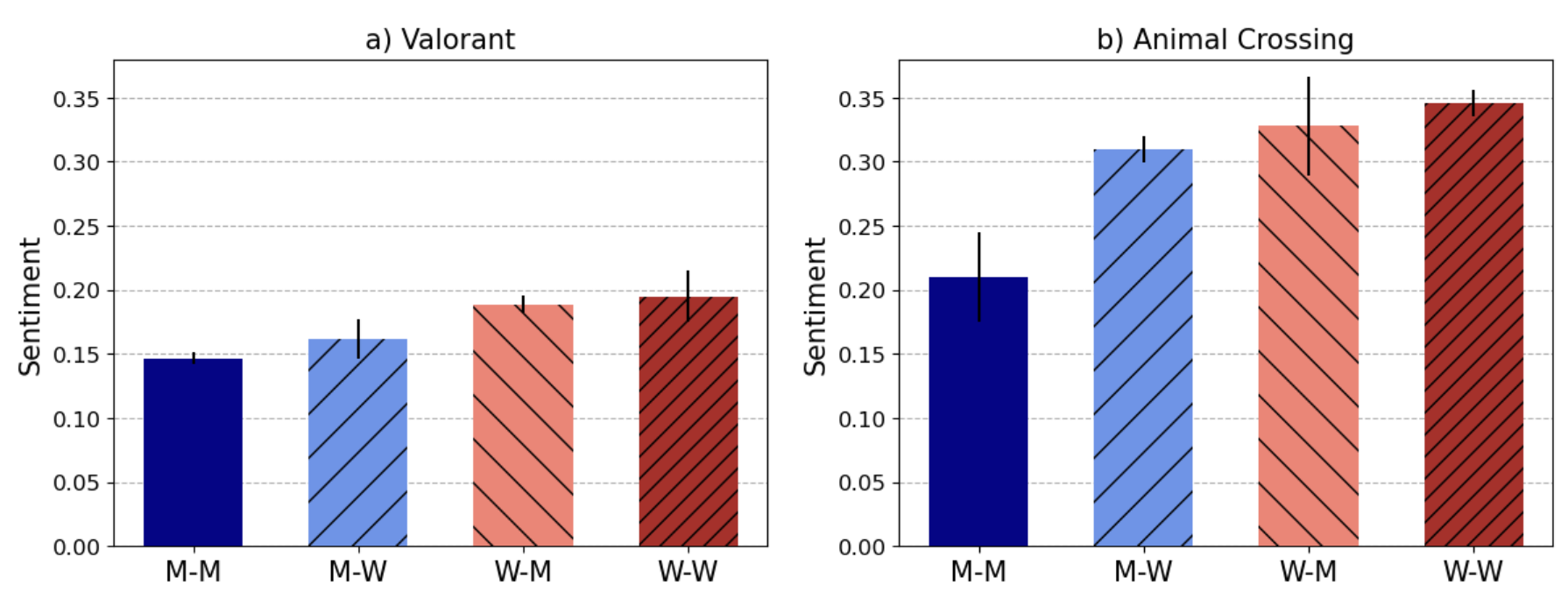}
  \caption{Mean Sentiment Measures of Gendered Collaborations in a) Valorant and b) Animal Crossing, by Collaboration Pair with Man-Man (Navy), Man-Woman (Light Blue), Woman-Man (Salmon), and Woman-Woman (Red).}
  \Description{Bar graph showing sentiment scores of gendered collaborations for a) Valorant and b) Animal Crossing. Sentiment scores from 0 to 0.35 are on the y-axis, and gendered pairs are on the x-axis. Sentiment patterns are similar across games, with W-W and W-M showing slightly higher positive sentiment than M-W and M-M. For Animal Crossing, sentiment is generally higher while exhibiting similar patterns to Valorant across gendered collaborations.}
  \label{fig:sentiment}
\end{figure}

\begin{figure}[H] 
  \centering
  \includegraphics[width=\linewidth]{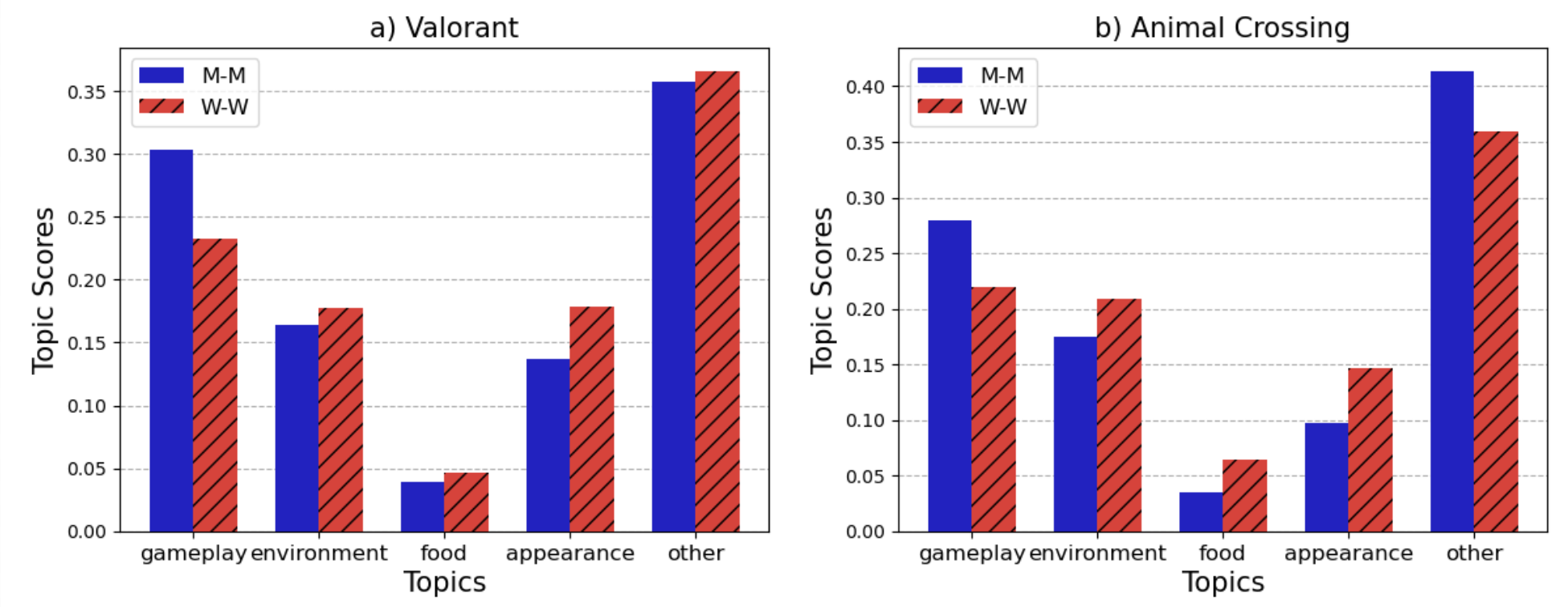}
  \caption{Topic Measures of Comment Discourse for a) Valorant and b) Animal Crossing, by Collaboration Pair with Man-Man (Blue) and Woman-Woman (Red).}
  \label{fig:topics}
  \Description{Bar graphs of topic scores in a) Valorant and b) Animal Crossing based on "Man-Man" (blue) and "Woman-Woman" (red) collaborations. Topic scores from 0 to 0.35 are on the y-axis, and topic categories are on the x-axis. Man-Man and Woman-Woman bars are placed side by side for each of the five topic categories: gameplay, environment, food, appearance, and other.} 
\end{figure}

Collaboration dynamics and synergies across genres vary greatly on the supply side. However, the demand side paints a different picture. Figure~\ref{fig:sentiment} shows the average sentiment of comments across a) \textit{Valorant} and b) \textit{Animal Crossing}. Figure~\ref{fig:topics} shows the proportion of comments that discuss particular topics in a) \textit{Valorant} and b) \textit{Animal Crossing} comment sections. \textit{Dead by Daylight} was redacted due to a lack of W-W collaborations.

Immediately, in Figure~\ref{fig:sentiment}, we observe women-led collaborations yield more positivity than men-led collaborations. The more woman-dominated a collaboration is, the more positive the resulting audience comments become. Moreover, the average sentiment value of non-collaborating video comments for \textit{Animal Crossing} is 0.287. However, the meaning comment sentiment of M-M collaborations is 0.210, meaning M-M collaborations have the potential to decrease the sentiment value. The average sentiment of \textit{Valorant}'s non-collaborating video comments was 0.137, so all collaborations increased the sentiment value.

Sentiment analysis is limited in that it does not reveal underlying context. We note these specific dimensions (streaming environment, food, person’s appearance) as they are well-known peripheral elements in gaming content~\cite{sjoblom2019ingredients, pollack2021twitch, ruberg2021livestreaming, nakandala2017gendered}. However, a more detailed typology should be created as well. Figure~\ref{fig:topics} shows the results of topic analysis across the categories: gameplay, environment, food, appearance, and others. Between M-M (blue) and W-W (red) collaborations, M-M collaborations yield significantly more attention on gameplay, whereas W-W collaborations yield significantly more discussion about appearance in both games. W-W collaborations  yield slightly more discussion about a creator's streaming environment and food. 

In sum, M-M collaborations yield more discourse about technical aspects of the game, whereas W-W collaborations yield more about peripherals.

\subsection{More Content Collaborations Does Not Indicate a More Connected Community}

\begin{figure}[h] 
 \centering
  \includegraphics[width=0.6\linewidth]{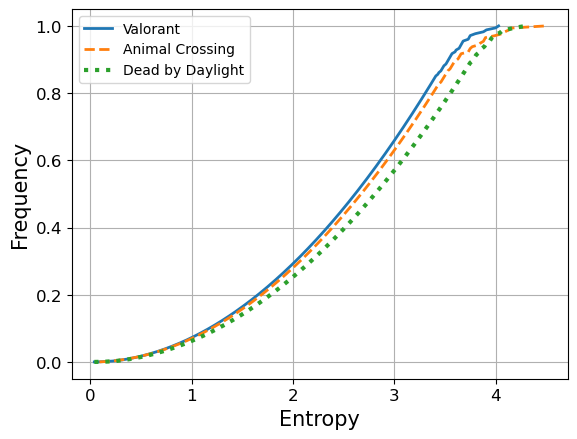}
  \caption{Entropy Measures of Commenters, by Games with Valorant (Blue), Animal Crossing (Orange), and Dead by Daylight (Green)}
  \Description{Line graph showing the frequency distribution of entropy for "Valorant" (solid blue line), "Animal Crossing" (dashed orange line), and "Dead by Daylight" (dotted green line). The x-axis represents entropy values ranging from 0 to 4.5, while the y-axis shows the cumulative frequency, ranging from 0 to 1.}
  \label{fig:entropy}
\end{figure}

Despite certain communities having more collaborations between YouTubers, that may not impact the connectivity of its community. Although \textit{Valorant} has the highest percentage of collaborating videos, viewers that comment on videos may tend to be "loyal" to one streamer rather than engaging cross channels. On the other hand, \textit{Dead by Daylight}, which had the lowest proportion of collaborating videos, has much more cross-engagement across channels. 

We quantify this through entropy analysis. Figure~\ref{fig:entropy} shows the cumulative density function of entropy scores amongst the commenters, with x-axis as the entropy and y-axis as the frequency. For \textit{Valorant}, the S-curve shifting more left indicates that the distribution of commenter interactions across channels is less diverse and more concentrated. \textit{Valorant} viewers tend to focus their attention on a few popular channels, perhaps due to the competitive nature of the game, where top players and influencers are seen as authoritative sources on strategies, gameplay, and entertainment. 

\textit{Animal Crossing} and \textit{Dead by Daylight} having entropy curves that are skewed right suggests a higher level of diversity in commenter interactions across YouTube channels. \textit{Animal Crossing}'s more creative and community-oriented gameplay likely encourages a broader range of content creators to attract attention from viewers. \textit{Dead by Daylight} being an asymmetric multiplayer horror game where players assume distinct roles--survivor or killer--creates a variety of gameplay experiences and strategies, which is reflected in content creation. Channels often specialize in different aspects of the game, such as survivor gameplay, killer gameplay, or specific characters, leading to a diverse content output~\cite{pcgamer}. As a result, viewers might spread their attention more evenly across channels that offer unique perspectives, leading to higher entropy and a right-skewed S-curve.

\section{Discussion}

In this study, we investigated inequalities in collaborative streaming from the supply and demand of attention. We chose three games, \textit{Valorant}, \textit{Animal Crossing}, and \textit{Dead by Daylight}, for their gendered ecosystems as comparative case studies. Additionally, we integrated Shapley value analysis for the study of collaborative synergy.

\subsection{Supply Side Dictated by Game Affordances and Demographics}

Past Human-Computer Interaction research has shown the overall patterns of online content collaboration outcomes. Building on this body of work, the current study provides further insights into which collaborative ties generate the most synergy. 

Through dyadic analysis of the supply-side, we find varying dominant ties for each game: man-man for \textit{Valorant}, woman-woman for \textit{Animal Crossing}, and man-woman for \textit{Dead by Daylight}. This does correlate with the demographics and audience makeup of the games and their known genres: \textit{Valorant} is predominantly male, \textit{Animal Crossing} is predominantly female. As such, our hypothesis was half correct. While there was a dominant dyad for each game that corresponded to the “gender” of the genre, out-group dyads can become dominant in the case of more gender-neutral genres (i.e. horror) or be potentially moderated by sexual orientation. 

Drawing from principles from evolutionary game theory, individuals are more motivated to cooperate when they anticipate rewards for cooperation, predict penalties for non-cooperation, or expect future interactions~\cite{van2011shadow, axelrod1981evolution}. In video game content collaborations, players are more incentivized to collaborate with those who maximize their potential gain. This becomes an issue when there is asymmetry between in-group and out-group collaborations. While in-group members enjoy the advantages of playing with a stable group of peers and also foster more in-group synergy, it comes at the cost of excluding others from valuable collaboration opportunities such as increased visibility and creative potential. Ultimately, distinct patterns of synergies will yield behavioral implications for exclusion and marginalization, leading to a cycle of reinforcing certain demographics of players in relation to games and game genres.

\subsection{Demand Side Dictated by Audience Gender Biases} 

On the demand side, regardless of which gender dominates the top channels in the game and the variation in game genres, audience responses towards different gendered collaborations are congruent in the communities that were investigated regardless of whether there was a dominant gender. 

Positive sentiment towards women-dominated collaborations indicates a more supportive community for female creators. Conversely, collaborations dominated by men creators tend to generate lower sentiment and more toxic audience engagement. Here, we build upon previous studies that have shown that collaborations tend to increase sentiment of audience response~\cite{irawan2023viewer}. We demonstrate that not only is the increase of sentiment impacted by genre of content, but also depends on the collaboration paradigms (i.e. gender). Further, in certain types of collaborations, such as man-man collaborations in \textit{Animal Crossing}, collaboration actually lowers the audience sentiment. 

When it comes to topics of discourse below videos, for woman-woman collaborations, there is a significantly greater proportion of comments directed to peripherals (i.e. appearance, ambience, and food) than actual game play. These confirm known gender biases within the streaming community, where, despite receiving more positive comments, women still experience objectification and a tendency of their audience  to focus on their appearance rather than on their gameplay and content~\cite{su2011virtual,nardi2010my,nakandala2017gendered}. This remains true women-led collaborations regardless of the genre of game. 

As such, our most crucial finding is as follows: we find asymmetric patterns amongst the supply-side, but symmetric patterns on the demand side. That is, the dominant gendered pair may vary greatly based on genre. Despite variations in dominant gendered collaborations, audience discourse and reactions to different gendered collaborations appear to be consistent. The variation in collaboration dynamics across gaming communities demonstrates that the structure, norms, and practices of collaborations are influenced by the specific affordances and characteristics of each game. 

\subsection{Limitations and Future Work}

Our study faces a few limitations. Future research could benefit from having a larger sample of games or genres than we used in order to better understand how gender dynamics manifest across collaborations in different communities. A larger dataset of videos within each game or genre could also provide more robust insights. For instance, there was a lack of representation of creators who identify outside of the gender binary within the top creators of each game analyzed. Future research can strive to include a more diverse range of gender identities among creators to capture a broader spectrum of collaboration dynamics. Another limitation is the challenge of defining which streamers are most relevant to include in the dataset; refining the criteria for selecting influential streamers could enhance the accuracy and relevance of future analyses. Finally, on extending Shapley values to greater sets of channel collaborations, the issue that arises is the potential for data sparsity. Dual collaborations are much more common than multi-channel collaborations. However, for certain genres of content, the option of extending this analysis exists.

\section{Conclusion}

In this study, we analyzed the collaborative dynamics of gaming content across game genres. Our results indicate that on the demand side, participation in collaboration, ease of collaboration, and synergy of collaboration results from game affordances and player demographics of the particular genre of game. On the supply side, patterns of audience reactions to gendered collaborations remain consistent across games, suggesting that these responses are influenced by prevailing gender biases that span across gaming communities. 

In sum, our work makes several contributions to research on creator collaborations and gender inequity in the gaming space. Empirically, we extend the understanding of genre affordances, how collaborations benefit each collaborator, and how interactions may affect the viewers’ engagement. Methodologically, we present a method to quantify the benefit of collaborations drawing from a canonical framework in game theory--Shapley values--which offers an extension to quantify synergies of collaborations.

Online collaboration differs substantially depending on the contributors' gender and status in the community. This should be taken into account when analyzing collaborative success, and may be insightful to communities facing gender gap and stagnation in collaboration participation levels in certain demographics. Understanding the dynamics of online collaborations, particularly through the lens of gender, offers important insights into how digital networks operate, how content creators navigate these spaces, and how communication patterns reflect and perpetuate existing inequalities.

\bibliographystyle{ACM-Reference-Format}
\bibliography{author}


\end{document}